\documentclass[12pt]{iopart}

\usepackage[square,numbers]{natbib}
\usepackage{iopams}
\usepackage{amssymb}
\usepackage{xcolor}
\usepackage{bm}
\usepackage{graphicx}
\usepackage{datetime}
\usepackage{silence}
\usepackage{subcaption}

\renewcommand{\vec}[1]{\mathbf{#1}}

\newcommand{\epseff}{\epsilon_\mathrm{eff}}
\newcommand{\species}{s}
\newcommand{\fluxLabel}{r}
\newcommand{\normFluxLabel}{\rho}
\newcommand{\rotTrans}{\iota}
\newcommand{\coll}{\nu}
\newcommand{\speed}{v}
\newcommand{\radE}{\vec{E}_\fluxLabel}
\newcommand{\absE}{E_\fluxLabel}
\newcommand{\absAbsE}{\left|\absE\right|}
\newcommand{\normE}{v_{E}^{*}}
\newcommand{\normColl}{\coll^{*}}
\newcommand{\transpCoeffNoSuper}[3]{#1_{#2 #3}}
\newcommand{\transpCoeffSuper}[4]{\transpCoeffNoSuper{#1}{#2}{#3}^{#4}}
\newcommand{\Lij}[3]{\transpCoeffSuper{L}{#1}{#2}{#3}}

\newcommand{\DijS}[2]{\transpCoeffSuper{D}{#1}{#2}{*}}
\newcommand{\Loo}[1]{\Lij{1}{1}{#1}}
\newcommand{\DooS}[0]{\DijS{1}{1}}
\newcommand{\eiRat}{\Lij{1}{1}{e} / \Lij{1}{1}{i}}

\newcommand{\tauE}{\tau_{E}}
\newcommand{\issof}{\tauE^{\mathrm{ISS04}}}

\newdateformat{monthyeardate}{%
  \monthname[\THEMONTH] \THEYEAR}

\begin{document}

\title[Direct optimization of neoclassical ion transport]{Direct optimization of neoclassical ion transport in stellarator reactors}

\author{B.{\ }F.{\ }Lee$^{1}$, S.{\ }A.{\ }Lazerson$^2$, H.{\ }M.{\ }Smith$^3$, C.{\ }D.{\ }Beidler$^3$, and N.{\ }A.{\ }Pablant$^4$}

\address{$^1$ Columbia University, New York, New York, U.S.A.}
\address{$^2$ Gauss Fusion GmbH, Garching bei M\"unchen, Germany}
\address{$^3$ Max Planck Institute for Plasma Physics, Greifswald, Germany}
\address{$^4$ Princeton Plasma Physics Laboratory, Princeton, New Jersey, U.S.A.}
\eads{\mailto{brandon.lee@columbia.edu}, \mailto{samuel.lazerson@gauss-fusion.com}, \mailto{hakan.smith@ipp.mpg.de}, \mailto{craig.beidler@ipp.mpg.de}, \mailto{npablant@pppl.gov}}
\vspace{10pt}
\begin{indented}
\item[]August 2024
\end{indented}
\begin{abstract}
    We directly optimize stellarator neoclassical ion transport while holding neoclassical electron transport at a moderate level, creating a scenario favorable for impurity expulsion and retaining good ion confinement. Traditional neoclassical stellarator optimization has focused on minimizing $\epseff$, the geometric factor that characterizes the amount of radial transport due to particles in the $1/\coll$ regime. Under expected reactor-relevant conditions, core electrons will be in the $1/\coll$ regime and core fuel ions will be in the $\sqrt{\coll}$ regime. Traditional optimizations thus minimize electron transport and rely on the radial electric field ($\absE$) that develops to confine the ions. This often results in an inward-pointing $\absE$ that drives high-$Z$ impurities into the core, which may be troublesome in future reactors. In this work, we increase the ratio of the thermal transport coefficients $\eiRat$, which previous research has shown can create an outward-pointing $\absE$. This effect is very beneficial for impurity expulsion. We obtain self-consistent density, temperature, and $\absE$ profiles at reactor-relevant conditions for an optimized equilibrium. This equilibrium is expected to enjoy significantly improved impurity transport properties.
\end{abstract}


\submitto{\NF}

\section{Introduction}

Traditional neoclassical stellarator optimizations have focused on minimizing the geometric factor $\epseff$, which characterizes the particle and heat transport of the $1/\coll$ regime \cite{nemov_evaluation_1999}. The $1/\coll$ regime is a subset of the long mean free path (lmfp) regime with zero $\radE \times \vec{B}$ drift velocity, $\normE \equiv \absAbsE/\speed B_{0}=0$, where $\radE = \absE(\fluxLabel) \nabla \fluxLabel$ is the radial electric field, $\fluxLabel$ is a flux label, $\speed$ is particle speed, and $B_0$ is a reference magnetic field strength. 
Stellarator reactors are expected to achieve temperatures in excess of $10~\mathrm{keV}$ in their plasma cores, resulting in the electrons and fuel ions exhibiting neoclassical transport behavior characteristic of the lmfp regime. Because the mass of the electrons is much lower than that of the ions, their characteristic thermal velocity is much higher. Thus we may reasonably approximate $\normE \approx 0$ for electrons in a reactor, indicating that their neoclassical transport behavior will be very near the ``ideal'' $1/\coll$ regime behavior characterized by $\epseff$. For the heavier fuel ions, on the other hand, $\normE \approx 0$ is a poor approximation, indicating that they will not be near the $1/\coll$ regime and $\epseff$ is not expected to accurately characterize their neoclassical transport behavior. Instead, we must consider two additional lmfp regimes which are visualized in figure~12 of reference~\cite{helander_theory_2018}: the $\coll$ regime at extremely low collisionalities and the $\sqrt{\coll}$ regime at more moderate collisionalities. 
Particle and heat fluxes are approximately proportional to $\absAbsE^{-2}$ for the former and $\absAbsE^{-3/2}$ for the latter \cite{beidler_benchmarking_2011}. 
Roughly speaking, $\radE \times \vec{B}$ rotation helps limit the radial excursions of particles from flux surfaces when particle transport behavior is not entirely dominated by collisional effects.
Since $\epseff$ characterizes transport levels when $\normE = 0$, lowering $\epseff$ lowers the ``ceiling'' on radial transport for populations in all lmfp regimes and thus reduces overall radial neoclassical transport. Reduced electron transport is the source of the overall transport reduction in low-$\epseff$ configurations, so an inward-pointing $\radE$ ($\absE < 0$) develops to prevent charge separation \cite{ho_neoclassical_1987}, or equivalently, to make the net radial current zero (the so-called ``ambipolarity condition''). Intuitively speaking, the radial electric field serves to ``hold in'' the fuel ions. The case when $\absE < 0$ is known as the ``ion root'' ambipolar electric field.

In design studies, stellarator reactors have commonly targeted $\epseff$ values in the range $0.1\%$ -- $1\%$ \cite{beidler_helias_2001, ku_physics_2008}, which is sufficiently small to prevent neoclassical energy losses from being an obstacle to ignition but not small enough to preclude other design objectives from being taken into account during the optimization process. Such reactors are expected to have strong ion root electric fields. This provides a thermodynamic force that drives impurities (e.g., partially-ionized tungsten atoms from the divertor and other plasma-facing components) toward the core. If techniques to minimize the outward turbulent electron and fuel ion transport in stellarators (see, for example, references~\cite{kim_optimization_2024, goodman_quasi-isodynamic_2024}) continue to advance, devices designed with these techniques will presumably have significantly reduced turbulent impurity transport as well. At some point, the ``pinch'' effect of the ion root electric field may be able to overpower the ``screening'' effect of turbulence. In this case, impurities would accumulate in the core and cause substantial radiative losses that would likely make continuous reactor operation difficult. The presence of a neoclassical impurity pinch has long been recognized in the literature \cite{rutherford_impurity_1974, w_vii-a_team_impurity_1985, hirsch_major_2008, garcia-regana_neoclassical_2013, mollen_impurities_2015}, and several remedies have been proposed. For instance, the temperature gradient of the fuel ions may help counteract the pinch associated with a small ion root electric field \cite{velasco_moderation_2017, helander_impurity_2017}, but this will likely not be effective if a strong ion root is present throughout the plasma. Furthermore, tangential electric fields may make this effect relatively unhelpful as they often enhance the impurity pinch caused by the radial electric field, especially for high-Z impurities \cite{garcia-regana_electrostatic_2017, buller_collisional_2018, calvo_stellarator_2018, mollen_flux-surface_2018}. Buller et al. attempted to directly optimize the tangential electric field (through the electric potential) to prevent inward impurity flux but were only able to reduce it \cite{buller_recent_2021}.

Another possible solution exists for the impurity transport problem: if $\absE > 0$ in some region of the plasma, this will provide a thermodynamic force that drives impurities (especially those with high $Z$) radially out of that region. When an outward-pointing radial electric field solves the ambipolarity condition, it is referred to as the ``electron root'' ambipolar electric field. Operating stellarators with an electron root was historically considered superior to ion root operation since the larger $\absAbsE$ characteristic of the electron root is favorable for confinement \cite{mynick_effect_1983, hastings_ambipolar_1985}. 
On the other hand, experimental realizations of the electron root typically involve substantial electron heating that causes the electron temperature to be much higher than the fuel ion temperature \cite{hirsch_major_2008, maasberg_neoclassical_2000, yokoyama_core_2007, klinger_performance_2017, pablant_core_2018} and/or hollow density profiles \cite{hirsch_major_2008, maasberg_neoclassical_2000, pablant_core_2018}. These situations are unlikely to be favorable or realizable in reactor scenarios due to the presence of turbulence, even if it is reduced through optimization. Furthermore, both situations contradict the desire for reactors to have strong collisional coupling (particularly in the core) such that plasma thermal energy efficiently drives fusion reactions.
This has lead to skepticism that stellarator reactors can feasibly operate with a steady-state electron root.

In this work, we seek to show that a steady-state electron root may be realistic in stellarator reactors.
Under the assumptions that transport processes are radially local and the drift kinetic equation can be linearized about a Maxwellian,
the neoclassical particle flux may be written as
\begin{equation}
    \Gamma_{\species}^{\mathrm{neo}} = -n_{\species} L_{11}^{\species} \left( \frac{n'_\species}{n_\species} - \frac{q_{\species} \absE}{T_\species} + \left( \frac{L_{12}^{\species}}{L_{11}^{\species}} - \frac{3}{2} \right) \frac{T'_\species}{T_\species} \right),
    \label{eq:flux}
\end{equation}
where $q_\species = Z_{\species} e$ (with $e$ being the elementary charge), $n_{\species} = n_{\species}(\fluxLabel)$, and $T_\species = T_{\species}(\fluxLabel)$ are the charge, density, and temperature of species $\species$, respectively, and the thermal transport coefficients are given by
\begin{equation}
    \Lij{i}{j}{\species} = \frac{1}{q_{\species}^{2} \rotTrans R_{0} B_{0}^{2}} \sqrt{\frac{\pi m_{\species} T_{\species}^{3}}{8}} \int_0^{\infty} dK_{\species} K_{\species}^{2} e^{-K_{\species}} \DijS{i}{j}(K_{\species}) h_{i} h_{j},
    \label{eq:Lij}
\end{equation}
with $\rotTrans$ the rotational transform on a given flux surface, $R_0$ a reference major radius, $m_\species$ the mass of species $\species$, $K_{\species} = \frac{m_{\species} v_{\species}^{2} / 2}{T_{\species}} = \left( \frac{v_{\species}}{v_{\species,\mathrm{thermal}}} \right)^{2}$, $\DijS{i}{j}$ the monoenergetic transport coefficient evaluated for a given $\normE$, $h_{1} = h_{3} = 1$, and $h_{2} = K_{\species}$ \cite{beidler_benchmarking_2011}.
While it may initially appear that $\eiRat \ll 1$ will always hold given the mass scaling in equation~\ref{eq:Lij},
it has been previously noted (to help explain the ``impurity hole'' phenomenon of the Large Helical Device \cite{yoshinuma_observation_2009, ida_observation_2009}) that sufficiently large $\eiRat$ will cause $\absE > 0$ \cite{velasco_moderation_2017}. 
Intuitively, this is because large $\eiRat$ signifies that radial electron transport is greater (in some sense) than radial ion transport, necessitating an outward-pointing radial electric field to maintain ambipolarity. Mathematically, raising $\eiRat$ increases the value of $\absE$ necessary to maintain ambipolarity when the density and temperature gradients are negative --- see equation~(3) of reference~\cite{velasco_moderation_2017}.
Optimized configurations have also recently been discovered by Beidler and coworkers with moderate values of $\epseff$ and extremely small $\DooS$ at fuel-ion-relevant $\normE$ \cite{beidler_reduction_2024},
leading to a large $\eiRat$ and therefore a strong electron root in the core under reactor-relevant conditions. The theoretical foundations of this phenomenon are explored in more detail in recent work by Helander et al.{\ }\cite{helander_optimised_2024}.

The objective of our work is to follow the suggestion of reference~\cite{beidler_reduction_2024} and optimize a stellarator magnetic configuration to increase $\eiRat$. We control $\Loo{e}$ through $\epseff$ and $\Loo{i}$ through an array of $\DooS$ evaluated at fuel-ion-relevant $\normE$ and $\normColl \equiv R_{0} \coll / \rotTrans \speed$ (normalized collisionality) values. 
Rather than driving $\epseff$ as low as possible and relying on the ion root that develops to confine the fuel ions, we directly minimize the fuel-ion-relevant $\DooS$ such that the fuel ions (rather than the electrons) become the rate-determining species. This encourages the development of an electron root to ``hold in'' the electrons, rather than an ion root to ``hold in'' the fuel ions. 
Monoenergetic transport coefficients should, in principle, provide a more general target than thermal transport coefficients because they are characteristic of a given magnetic field and do not depend on plasma profiles.
Our transport optimization is entirely neoclassical --- because turbulence is often the dominant transport mechanism in optimized stellarators such as Wendelstein 7-X \cite{carralero_experimental_2021}, the techniques presented here may only be useful when coupled with turbulence optimization techniques. 
We note that large $\absE'$ (and thus large $\radE \times \vec{B}$ shear), which is common in electron root scenarios, may help reduce radial turbulent fluxes \cite{terry_suppression_2000, kuczynski_self-consistent_2024}. 
Large $\absE'$, usually in the ion root region near the last closed flux surface, is known to allow access to improved confinement modes of operation somewhat similar to H-modes in tokamaks \cite{hirsch_major_2008, wagner_quarter-century_2007, estrada_sheared_2009}.

\section{Formulation}
\label{sec:formulation}

\begin{table}
    \centering
    \caption{Optimization objectives used to produce the example configuration. The neoclassical targets are evaluated on the $\normFluxLabel = \{0.25, 0.5, 0.75\}$ flux surfaces. The rotational transform is evaluated on the $s = \normFluxLabel^2 = \{0, 0.5, 1\}$ flux surfaces.}
    \begin{tabular}{|c|c|c|}
        \hline
        Objective & Target Value & Weight $\left( 1/\sigma \right)$ \\
        \hline
        $\epseff$ & $1\%$ & $10^{2}$ \\
        Fuel-Ion-Relevant $\DooS$ & 0 & $10^{2}$ \\
        Rotational Transform & $\{8.14 \times 10^{-1}, 8.70 \times 10^{-1}, 9.47 \times 10^{-1}\}$ & $10^{0}$ \\
        Average Boundary Elongation & 1.87 & $10^{-1}$ \\
        Aspect Ratio & 10.89 & $4 \times 10^{-2}$ \\
        Volume & $1755.76~\mathrm{m^{3}}$ & $10^{-2}$ \\
        \hline
    \end{tabular}
    \label{tab:opt}
\end{table}

STELLOPT \cite{lazerson_stellopt_2020} is the primary optimization software used in this work. STELLOPT attempts to optimize finite-$\beta$ MHD equilibria by minimizing an objective function of the form
\begin{equation}
	\chi^{2} = \sum_{i} \frac{\left( f_{i} - f_{i}^{\mathrm{target}} \right)^{2}}{\sigma_{i}^{2}},
	\label{eq:chi}
\end{equation}
where $f_{i}$ is any quantity that is callable by the STELLOPT routines, $f_{i}^{\mathrm{target}}$ is the desired value for $f_{i}$, and $\sigma_{i}$ is a user-specified inverse weight. VMEC \cite{hirshman_steepest-descent_1983} is used as the MHD equilibrium solver in all cases. Differential evolution \cite{storn_differential_1997} and a Garabedian boundary representation \cite{garabedian_quasi-axially_1998} are utilized to find global minima more reliably. We allow STELLOPT to modify all the boundary modes of the initial configuration to minimize the objective function.

\begin{figure}
    \centering
    \includegraphics[width=0.5\textwidth]{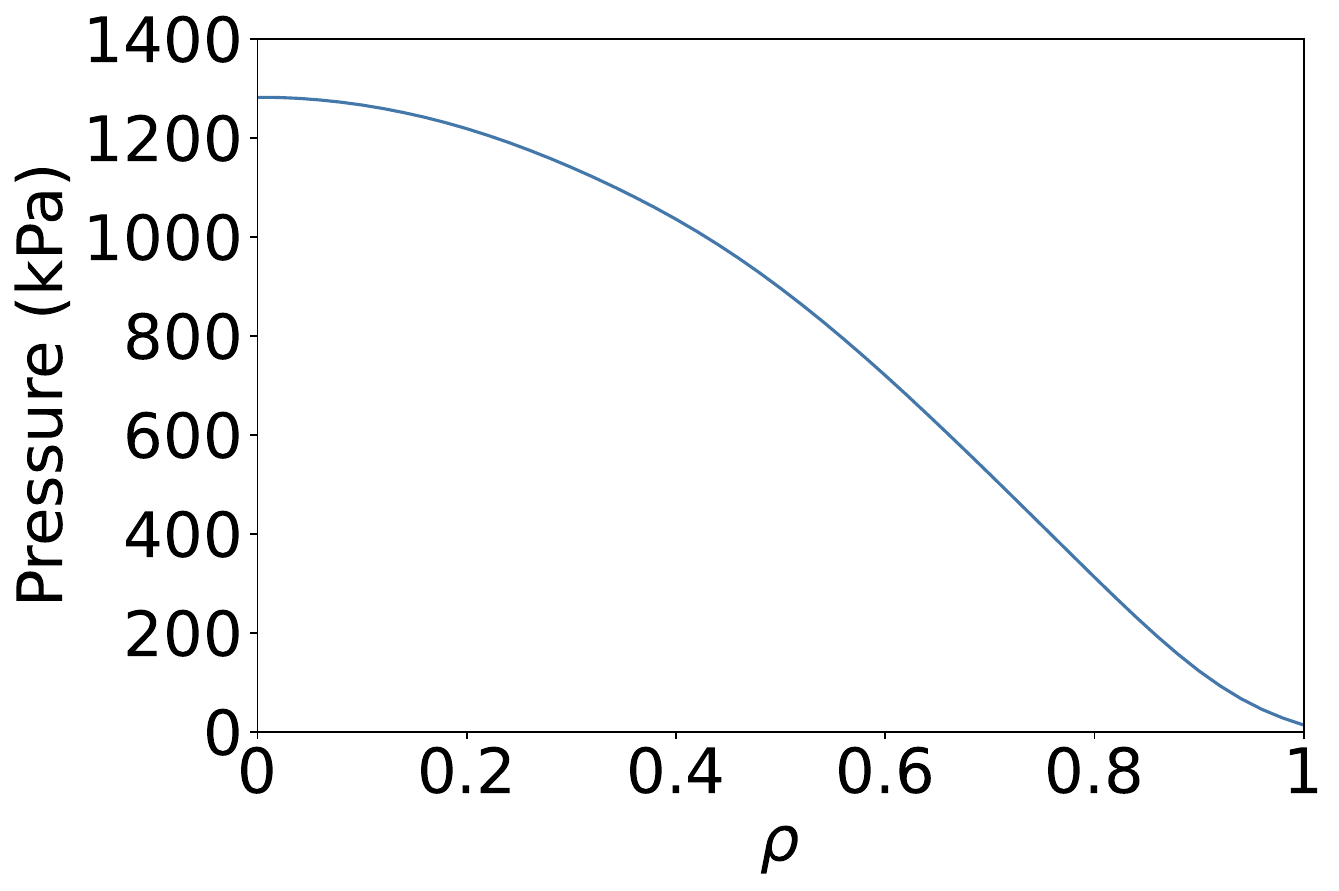}
    \caption{The pressure profile assumed during the finite-$\beta$ optimization of the example configuration.}
    \label{fig:pressure}
\end{figure}

The optimization begins from a fixed-boundary configuration based on the \mbox{W7-X} high-mirror configuration \cite{dinklage_magnetic_2018} 
whose volume and on-axis magnetic field strength have been scaled to the reactor-relevant values of roughly 1756~$\mathrm{m^3}$ and 5~T, respectively \cite{lazerson_simulating_2021}.
Stellarator symmetry and five field periods are assumed throughout. The initial value of $\beta$ is roughly $5\%$. The pressure profile used throughout the optimization can be seen in figure~\ref{fig:pressure}. Zero bootstrap current is assumed during the MHD calculations within the optimization loop. The core feature of the optimization is the minimization of an array of $\DooS$ with fuel-ion-relevant $\normE$ and $\normColl$. The $\DooS$ values are calculated by DKES \cite{hirshman_plasma_1986, van_rij_variational_1989}, which is coupled to STELLOPT such that it can be called within the optimization loop.
A $3 \times 3$ array of $\DooS$ values with $\absAbsE/\speed = \left\{ 5.00 \times 10^{-4}, 2.74 \times 10^{-3}, 1.50 \times 10^{-2} \right\}~\mathrm{T}$ and $\coll/\speed = \left\{ 2.50 \times 10^{-6}, 1.58 \times 10^{-5}, 1.00 \times 10^{-4} \right\}~\mathrm{m^{-1}}$ is used. (These quantities, rather than their normalized counterparts, are specified as targets in STELLOPT.) The exact values of $\absAbsE/\speed$ and $\coll/\speed$ seem relatively unimportant --- the upper and lower values for each quantity are chosen to be approximately reactor-relevant bounds, and the middle values are the logarithmic means of the upper and lower values. 
Notably, modifications of $B_0$, $R_0$, and $\rotTrans$ are relatively small in the optimization, so explicitly targeting $\absAbsE / \speed$ and $\coll / \speed$ rather than $\normE$ and $\normColl$ is acceptable.
In addition to $\DooS$, we target $\epseff = 1\%$ (calculated by NEO \cite{nemov_evaluation_1999}) during the optimization. These neoclassical objectives are calculated for the $\normFluxLabel = r/a = \{0.25, 0.5, 0.75\}$ flux surfaces.
We also include objectives for the average boundary elongation, aspect ratio, and volume to prevent strong boundary shaping or large changes in the size of the configuration. 
Finally, we target the rotational transform on the $s = \normFluxLabel^2 = \{0, 0.5, 1\}$ flux surfaces in an attempt to keep $\rotTrans' > 0$ as this simultaneously helps improve MHD \cite{mercier_necessary_1960} and neoclassical tearing mode \cite{hegna_stability_1994} stability. 
Note that we do not include the full Mercier criterion as an optimization objective.
The initial values of the MHD objectives are used as targets during the optimization --- in other words, we seek to hold MHD properties steady during the optimization while modifying neoclassical properties.
The neoclassical targets for $\epseff$ and $\DooS$ are the most important ingredients in the optimization, and the latter constitutes the novel approach of this work.
See table~\ref{tab:opt} for numerical details of the objective function.

Once optimized, the transport characteristics of the example configuration are analyzed using DKES and NTSS \cite{turkin_current_2006}.
NTSS performs flux-surface-averaged, time-dependent transport calculations (including momentum correction techniques \cite{maasberg_momentum_2009}) to determine self-consistent profiles for density, temperature, radial electric field, bootstrap current, and other quantities on 51 radial grid points for a given equilibrium. It should be emphasized that the equilibrium itself is not updated during the transport calculations, but NTSS calculates bootstrap current using the $\DijS{3}{1}$ database provided by DKES and updates the rotational transform profile by utilizing susceptance matrices \cite{strand_magnetic_2001}.
NTSS determines the full profile of the radial electric field by solving a diffusion-like differential equation with hysteretic behavior \cite{hastings_differential_1985} that minimizes the generalized heat production rate of the plasma \cite{shaing_stability_1984}.
This is necessary because the ambipolarity condition generally produces either one or three possible solutions for $\absE$ \cite{mynick_effect_1983, jaeger_neoclassical_1978, hastings_bifurcation_1986} and a discontinuous $\absE$ profile is nonphysical.
In the case of three solutions, one is the ion root, one is the electron root, and one is an unstable root that is not physically realized.
We choose the ``diffusion coefficient'' in the diffusion-like equation to be 2~$\mathrm{m^{2} s^{-1}}$ such that the transition region from the electron root to the ion root is roughly 5~cm wide, which is reasonable in light of global neoclassical simulations \cite{kuczynski_self-consistent_2024}.
Specifying a detailed startup scenario is beyond our scope, so we initialize NTSS runs by instructing the software to ``select'' the electron root solution if it exists. 
This situation is relevant for reactors that primarily utilize electron heating for startup.
When the configuration is loaded into NTSS, we rescale its volume to $1900~\mathrm{m^3}$ and its on-axis magnetic field strength to $6~\mathrm{T}$.
This allows us to achieve a steady-state alpha power of roughly 600~MW, which approximately corresponds to a 3~GWth reactor.
(We note that the electron root is expected to be more difficult to achieve as the magnetic field strength increases \cite{beidler_benchmarking_2011, kuczynski_self-consistent_2024}, which may need to be taken into account if our methods are used alongside those for turbulence optimization.) Deuterium and tritium density profiles are fixed with on-axis values for each in excess of $10^{19}$~$\mathrm{m^{-3}}$ and core gradients relatively flat in $\normFluxLabel$ --- see figure~\ref{fig:ntss} for details.
We assume there are no heavy impurities in the plasma. The temperature profiles for each species are initialized with roughly parabolic shapes in $\normFluxLabel$. The initial core electron temperature is set to $21.46~\mathrm{keV}$ and the initial core fuel ion temperature is set to $18.57~\mathrm{keV}$.
NTSS uses the D-T cross-section to calculate the birth rate of alpha particles, which for simplicity are assumed to slow down on the flux surface of their birth. This determines the helium density, and quasi-neutrality then determines electron density. 
A very simple turbulent transport model is included based on fits to \mbox{W7-AS} data. It takes the form \cite{ringler_confinement_1990}
\begin{equation}
	Q_{s}^{\mathrm{turb}} = -\chi_{\mathrm{turb}} n_{s} T'_{s},
    \label{eq:Qturb}
\end{equation}
\begin{equation}
	\Gamma_{s}^{\mathrm{turb}} = -D_{\mathrm{turb}} n'_{s},
    \label{eq:Gammaturb}
\end{equation}
where $Q_{s}^{\mathrm{turb}}$ is the radial energy flux of species $s$ due to turbulence.
Similarly to reference~\cite{beidler_reduction_2024}, we set
$\chi_\mathrm{turb} = 0.0065 (P/\mathrm{MW})^{3/4} / \left(n_{s} / 10^{20}~\mathrm{m^{-3}}\right)~\mathrm{m^{2} s^{-1}}$ and $D_\mathrm{turb} = 0.0003 (P/\mathrm{MW})^{3/4} / \left(n_{s} / 10^{20}~\mathrm{m^{-3}}\right)~\mathrm{m^{2} s^{-1}}$ where $P =  P_\alpha-P_\mathrm{Br}$, $P_\alpha$ is the power deposited by fast ions, and $P_{\mathrm{Br}}$ is the power radiated due to Bremsstrahlung. 
No external heating is considered.
NTSS steps forward in time until it converges on a steady-state, self-consistent solution for the plasma profiles. 
In this work, we seek a configuration that can realize an electron root and maintain it during steady-state operation.
(Configurations with insufficiently optimized ion confinement, such as scaled \mbox{W7-X} high-mirror, may be artificially seeded with an electron root during high-density operation but will quickly transition to an ion root.)

\section{Results}
\label{sec:results}
\begin{figure}
    \centering
    \begin{subfigure}[b]{0.35\textwidth}
    \includegraphics[width=\linewidth]{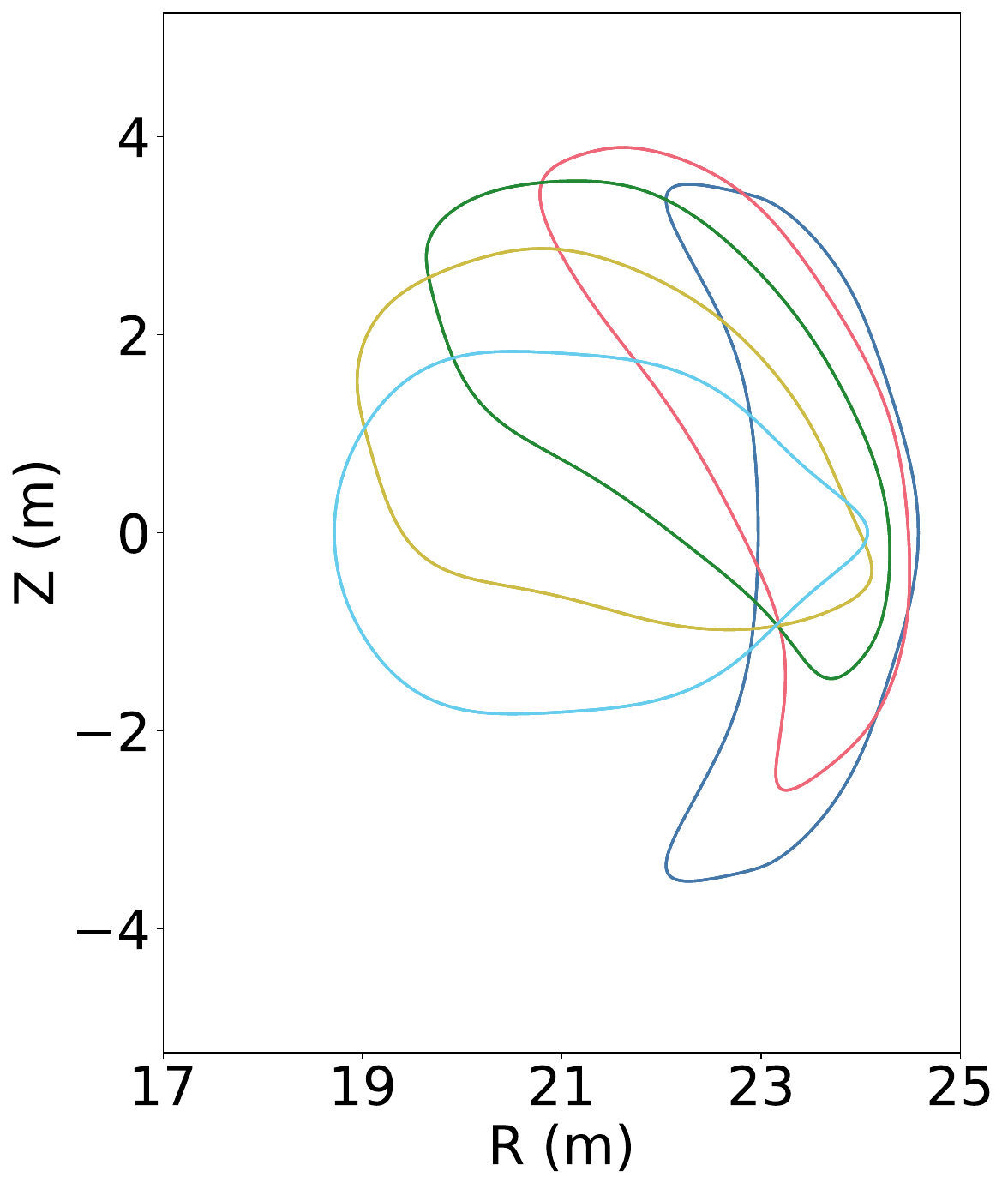}
    \caption{}
    \label{fig:config3_lpk}
    \end{subfigure}
    \begin{subfigure}[b]{0.35\textwidth}
    \includegraphics[width=\linewidth]{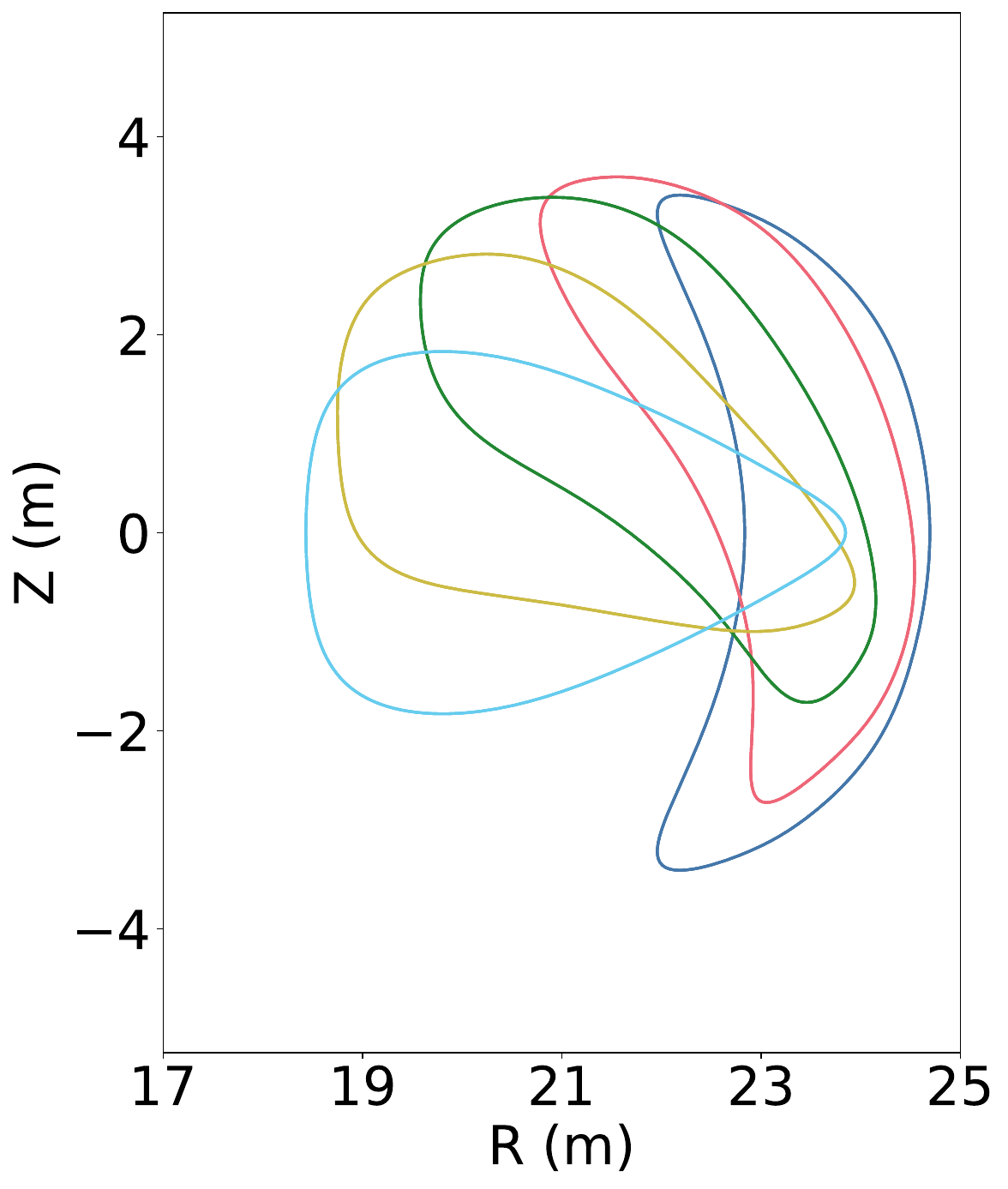}
    \caption{}
    \label{fig:w7x_lpk}
    \end{subfigure}
    \caption{Toroidal cross-sections of the last closed flux surface equally spaced in the toroidal angle across one half-field-period for (\subref{fig:config3_lpk}) the example configuration and (\subref{fig:w7x_lpk}) scaled \mbox{W7-X} high-mirror.}
    \label{fig:lpks}
\end{figure}

\begin{figure}
    \centering
    \begin{subfigure}[b]{0.43\textwidth}
    \includegraphics[width=\linewidth]{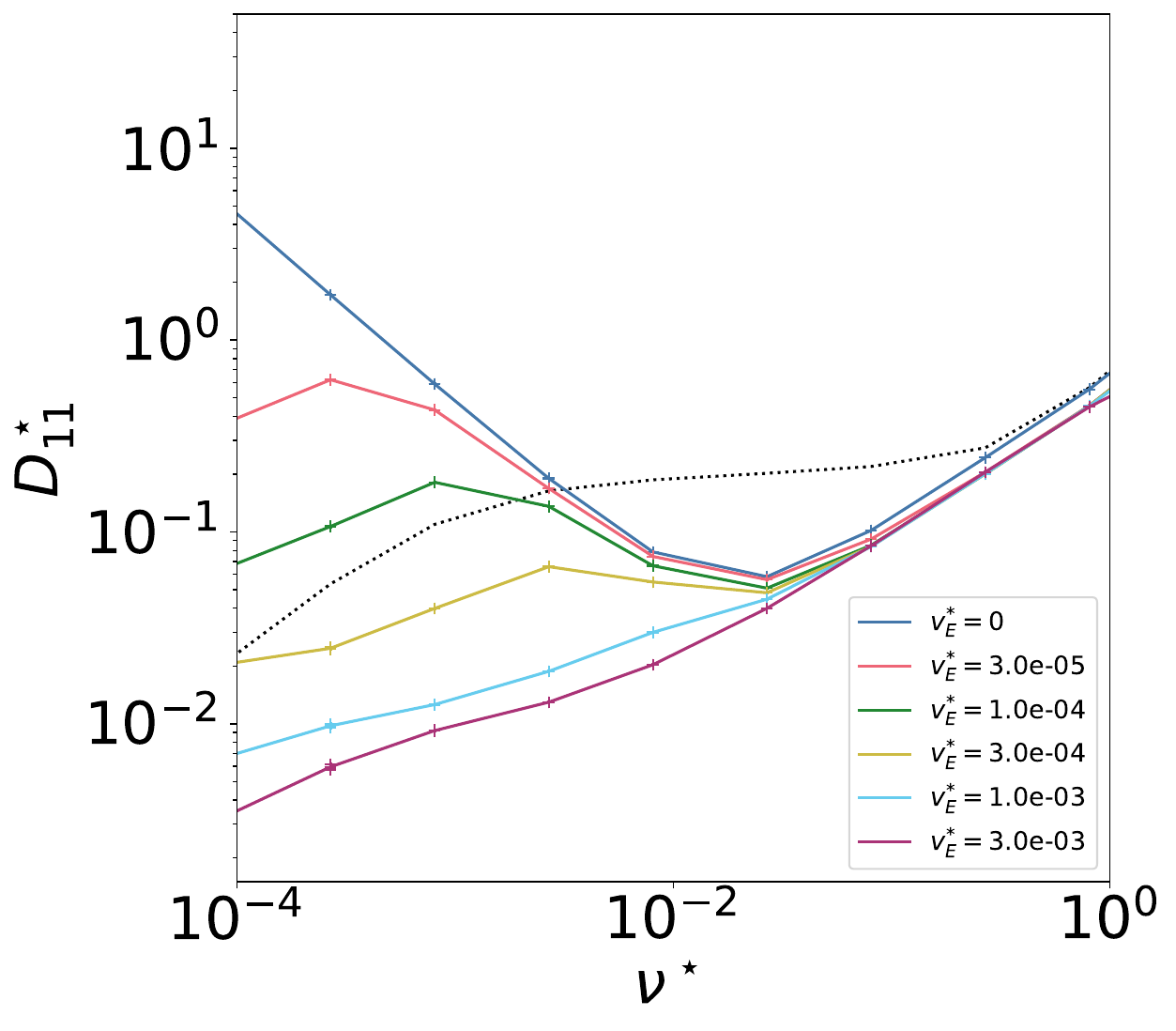}
    \caption{}
    \label{fig:config3_D11}
    \end{subfigure}
    \begin{subfigure}[b]{0.43\textwidth}
    \includegraphics[width=\linewidth]{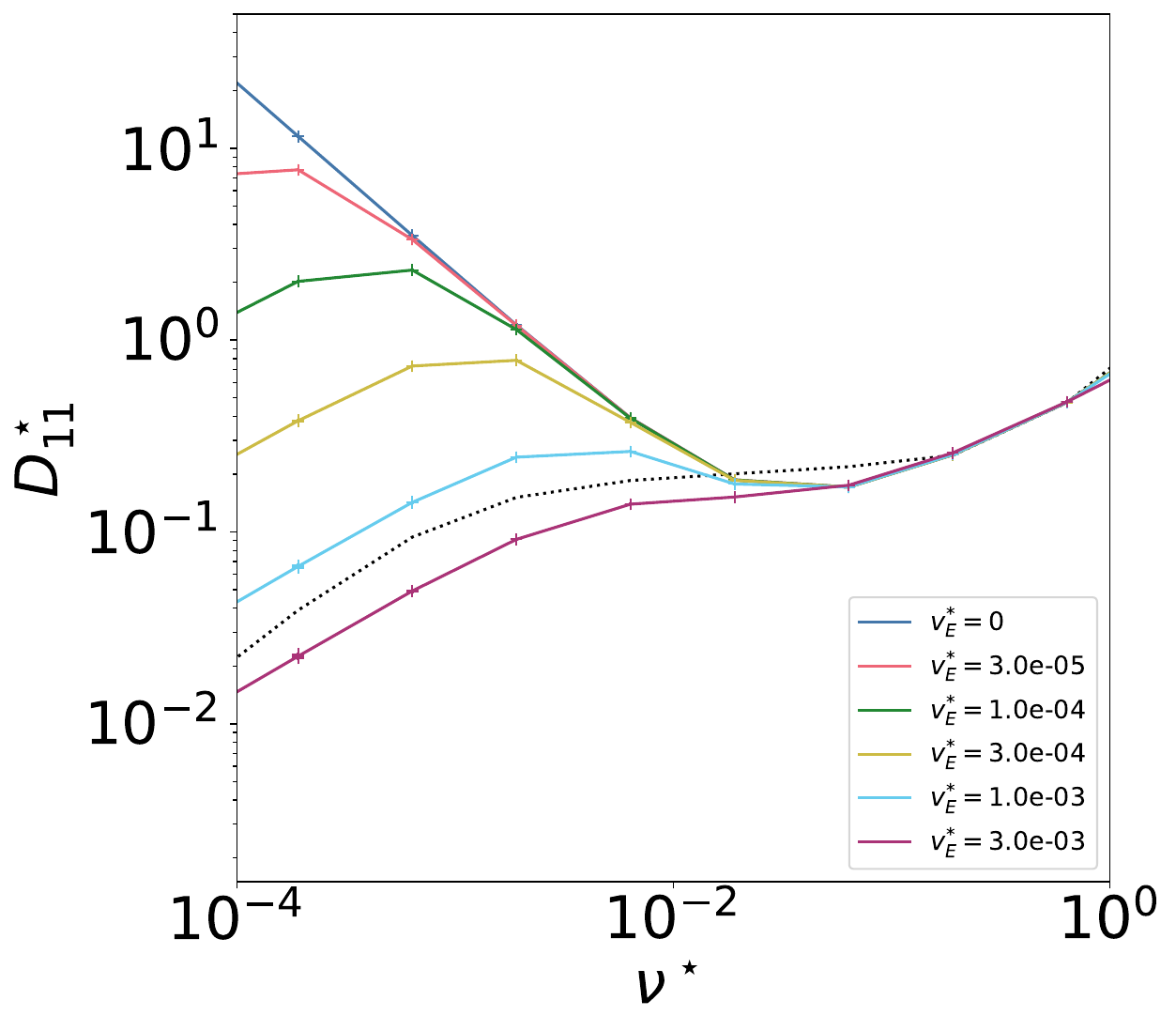}
    \caption{}
    \label{fig:w7x_D11}
    \end{subfigure}
    \caption{Scans of $\DooS$ versus $\normColl$ with various $\normE$ at $\normFluxLabel \approx 0.35$ for (\subref{fig:config3_D11}) the example configuration and (\subref{fig:w7x_D11}) \mbox{W7-X} high-mirror. The upper and lower variational bounds on the $\DooS$ values are calculated by DKES and bracketed by the ``$+$'' symbols, but these ``error bars'' are too small to see in most cases. The dotted lines indicate $\DooS$ for a tokamak with the same aspect ratio, rotational transform, and (average) elongation as the given configuration for the same $\normFluxLabel$. Note that the light blue and purple curves, roughly corresponding to fuel-ion-relevant $\normE$, are far more depressed in the example configuration than in \mbox{W7-X} high-mirror.}
    \label{fig:D11s}
\end{figure}

\begin{figure}
    \centering
    \includegraphics[width=0.5\textwidth]{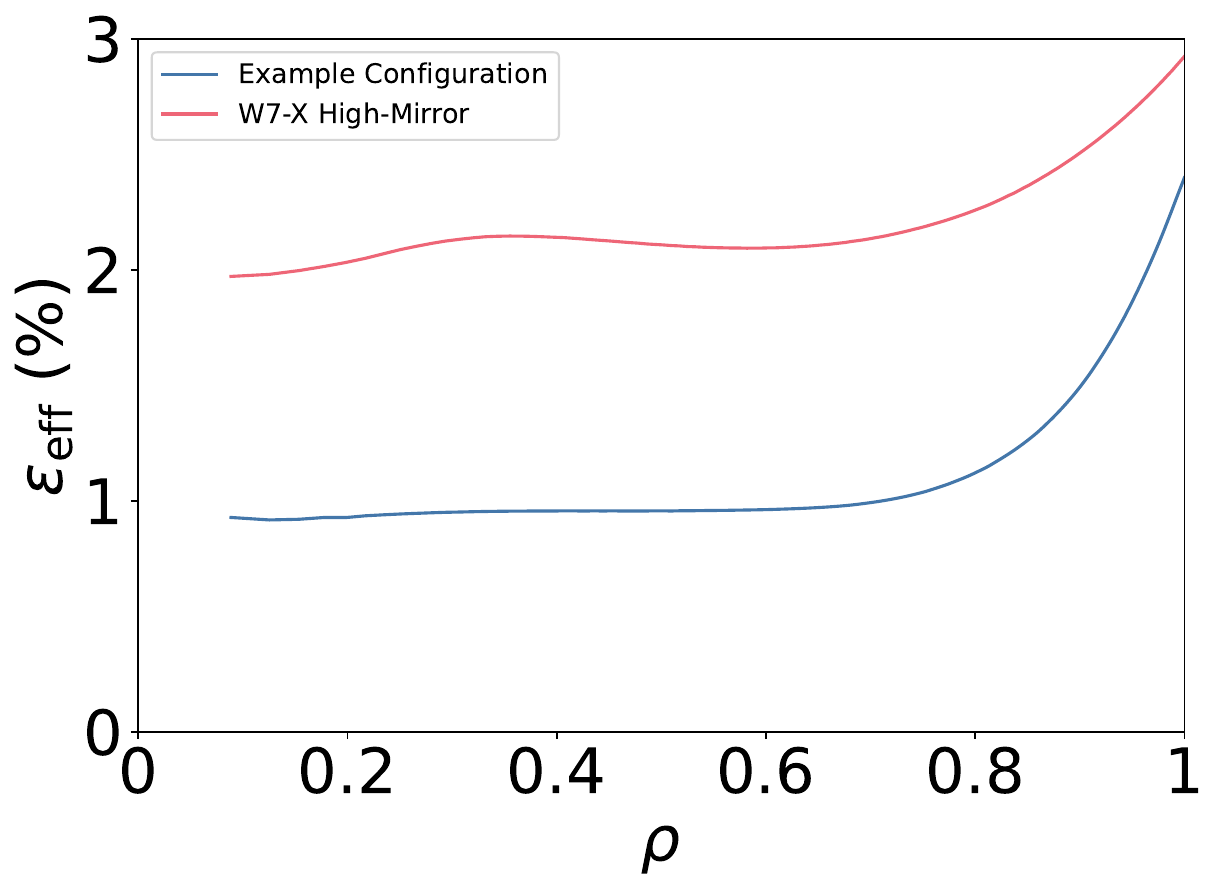}
    \caption{The geometric factor $\epseff$ for the example configuration and \mbox{W7-X} high-mirror. These calculations are performed on the optimized equilibria without taking effects from the NTSS modeling (such as modifications to bootstrap current, $\beta$, etc.) into account.}
    \label{fig:epseffs}
\end{figure}

\begin{figure}
    \centering
    \includegraphics[width=0.4\textwidth]{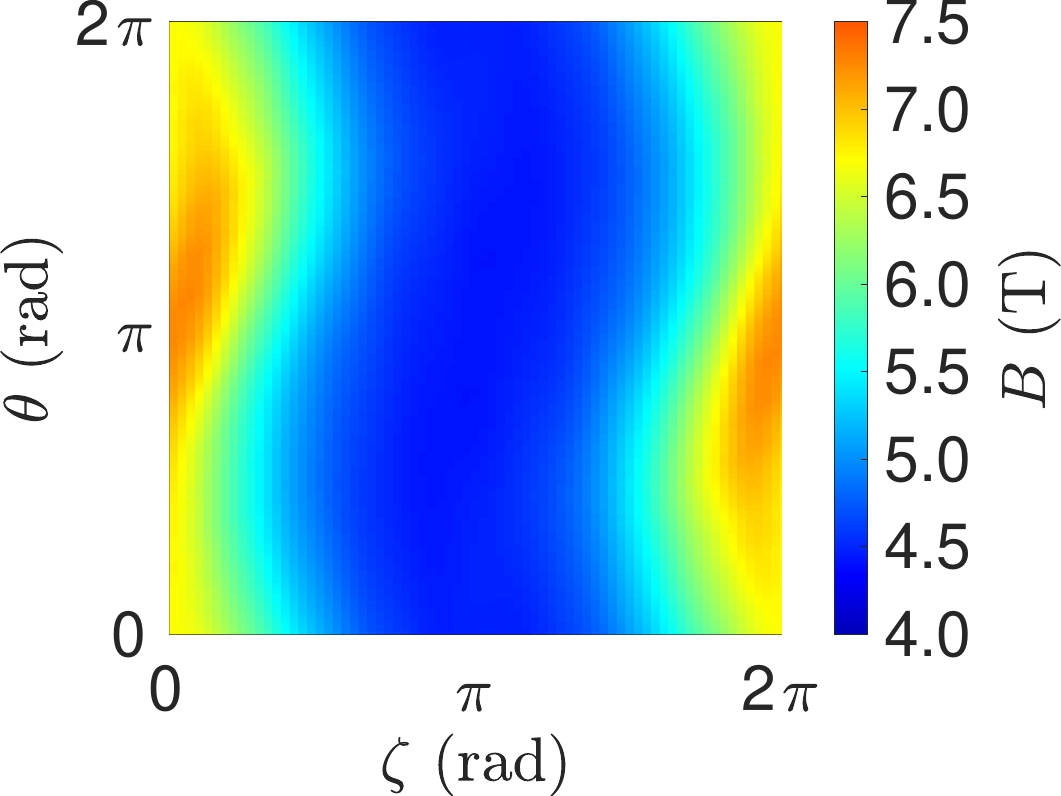}
    \caption{Magnetic field strength of the example configuration at mid-radius with an on-axis magnetic field strength of roughly $5.19~\mathrm{T}$. $\theta$ is the VMEC poloidal angle and $\zeta$ is the VMEC toroidal angle.}
    \label{fig:Bcont}
\end{figure}

\begin{figure}
    \centering
    \includegraphics[width=0.4\textwidth]{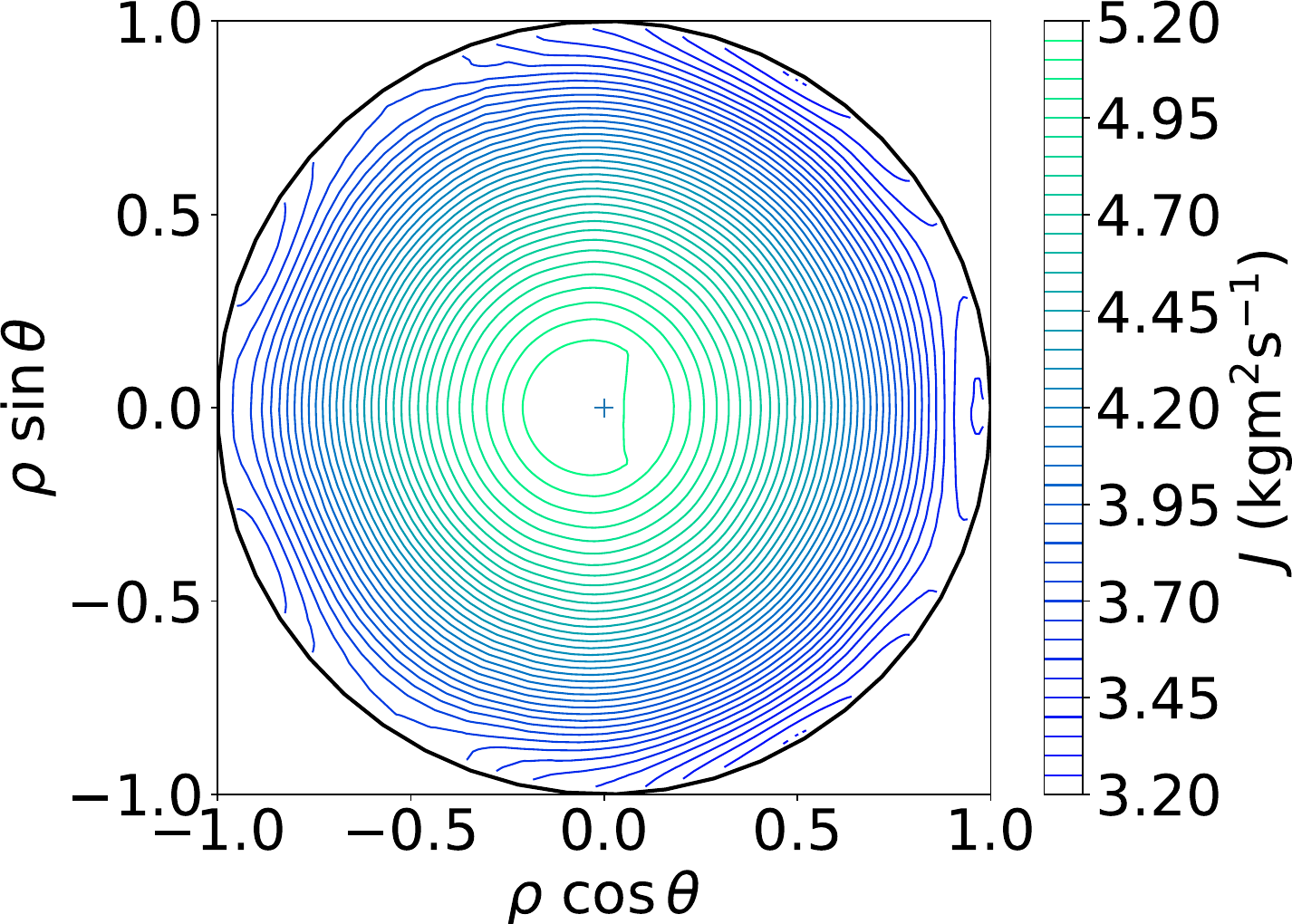}
    \caption{Contours of the second adiabatic invariant $J$ for the example configuration. This plot was produced using the methods of reference~\cite{faustin_fast_2016} with an on-axis magnetic field strength of roughly $5.19~\mathrm{T}$. $\theta$ is the VMEC poloidal angle.}
    \label{fig:Jcont}
\end{figure}

\begin{figure}
    \centering
    \begin{subfigure}[b]{0.43\textwidth}
    \includegraphics[width=\linewidth]{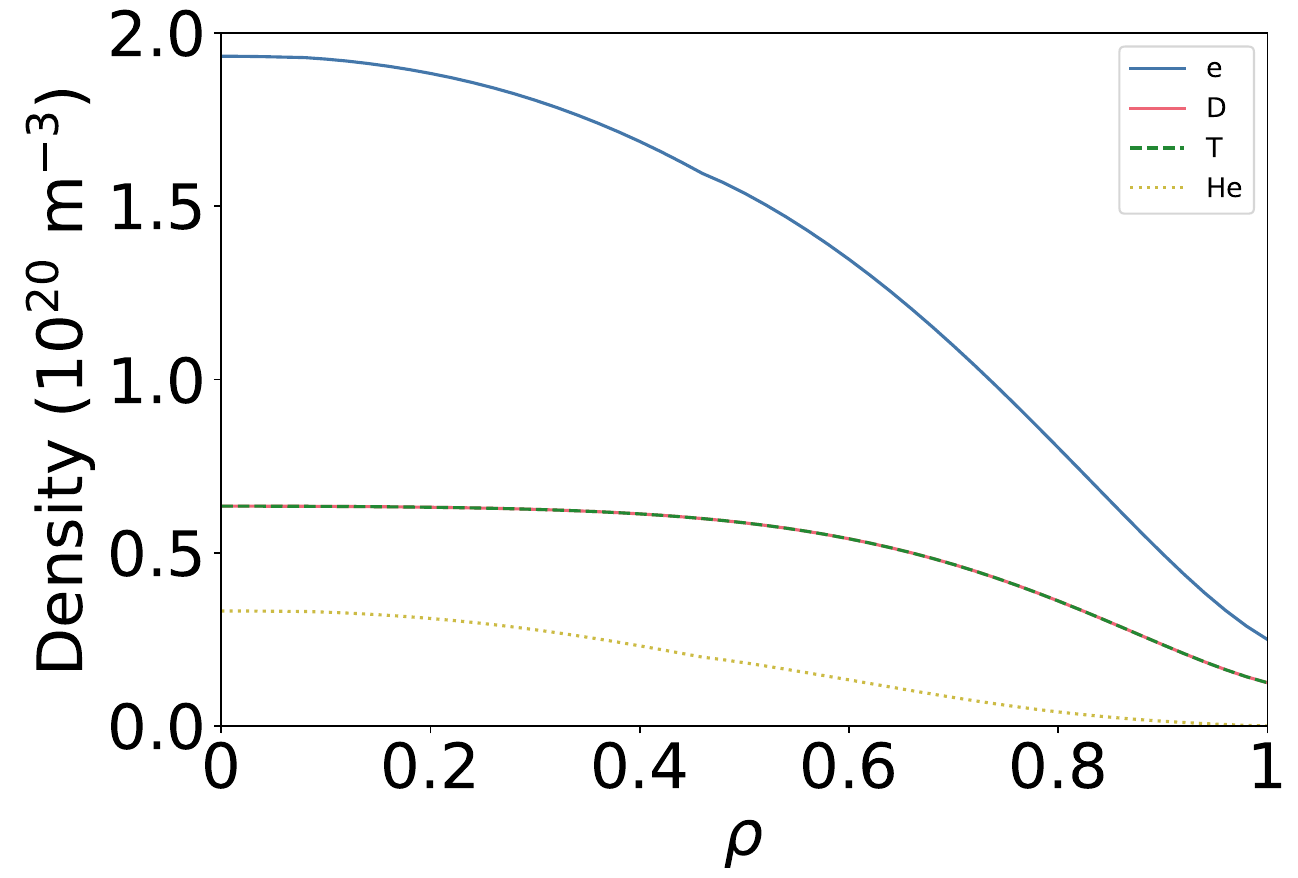}
    \caption{}
    \label{fig:lee3_n}
    \end{subfigure}
    \begin{subfigure}[b]{0.43\textwidth}
    \includegraphics[width=\linewidth]{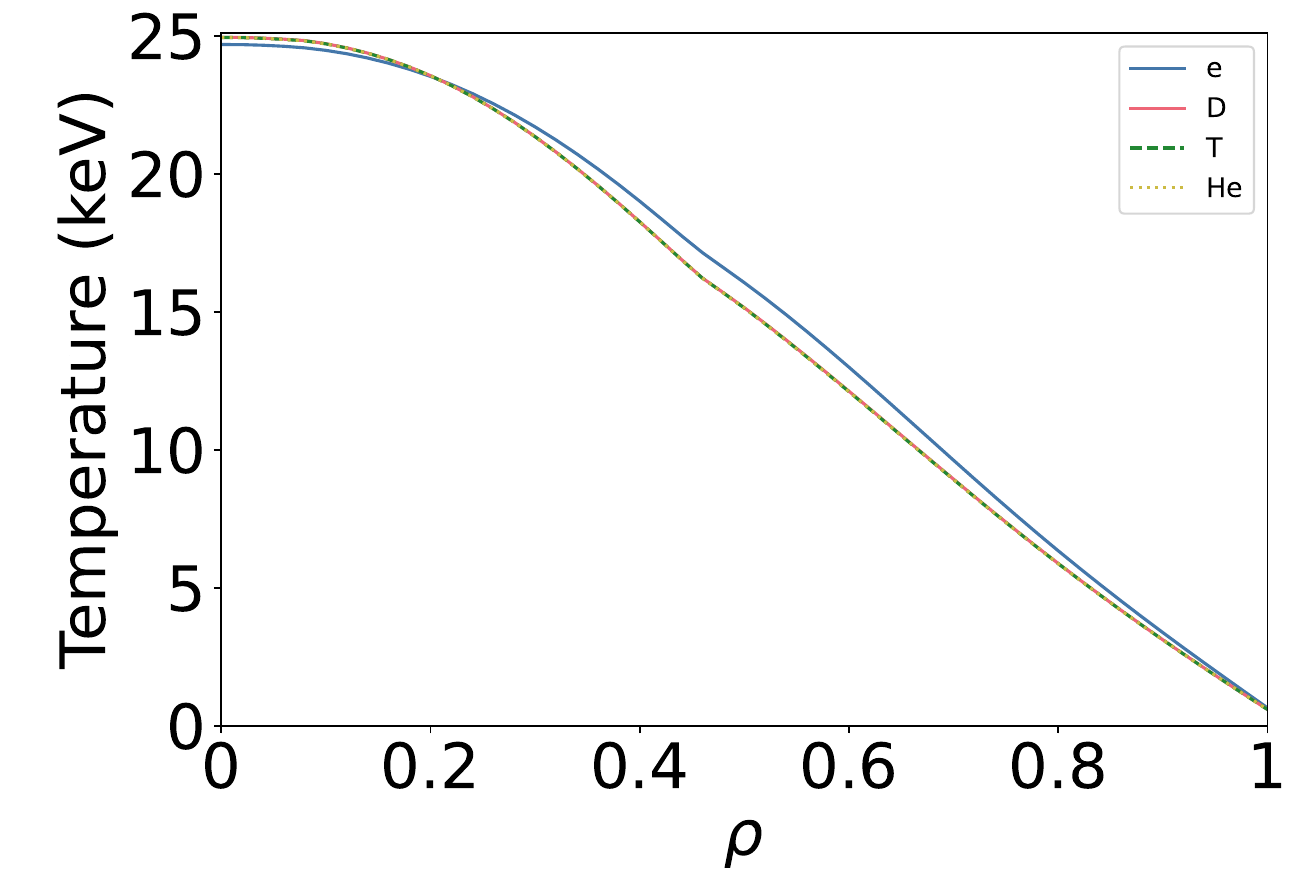}
    \caption{}
    \label{fig:lee3_T}
    \end{subfigure}
    \begin{subfigure}[b]{0.43\textwidth}
    \includegraphics[width=\linewidth]{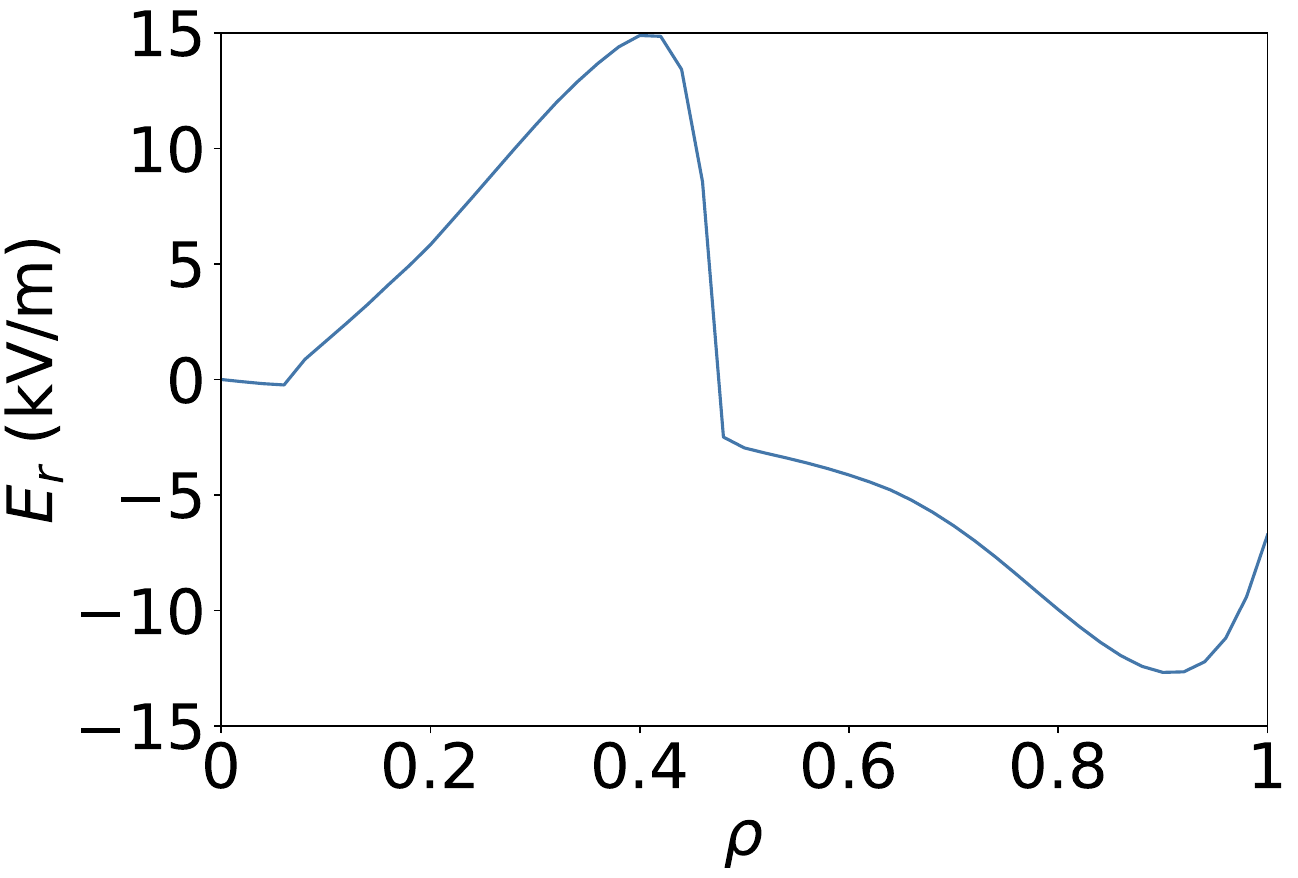}
    \caption{}
    \label{fig:Ers}
    \end{subfigure}
    \begin{subfigure}[b]{0.43\textwidth}
    \includegraphics[width=\linewidth]{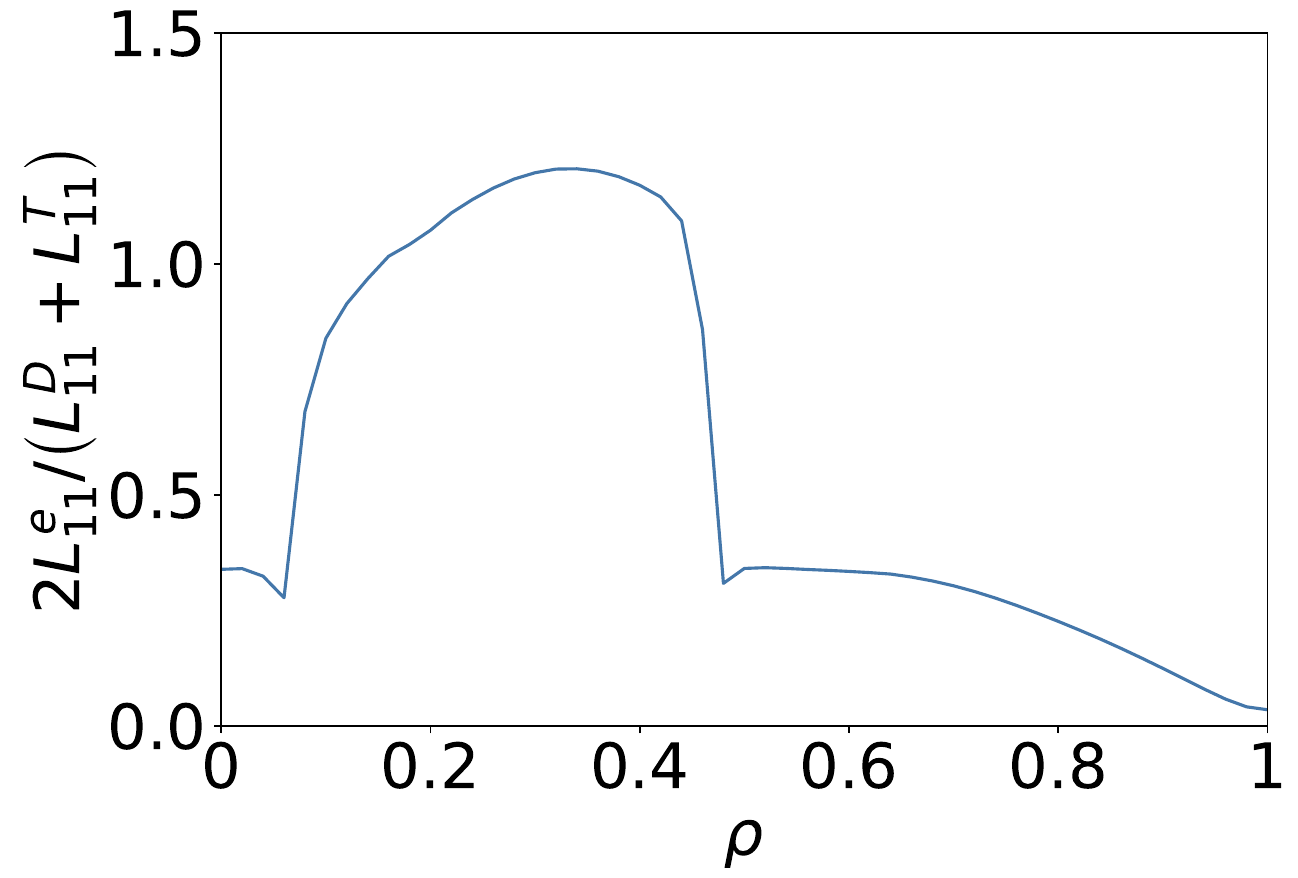}
    \caption{}
    \label{fig:L11Rats}
    \end{subfigure}
    \begin{subfigure}[b]{0.43\textwidth}
    \includegraphics[width=\linewidth]{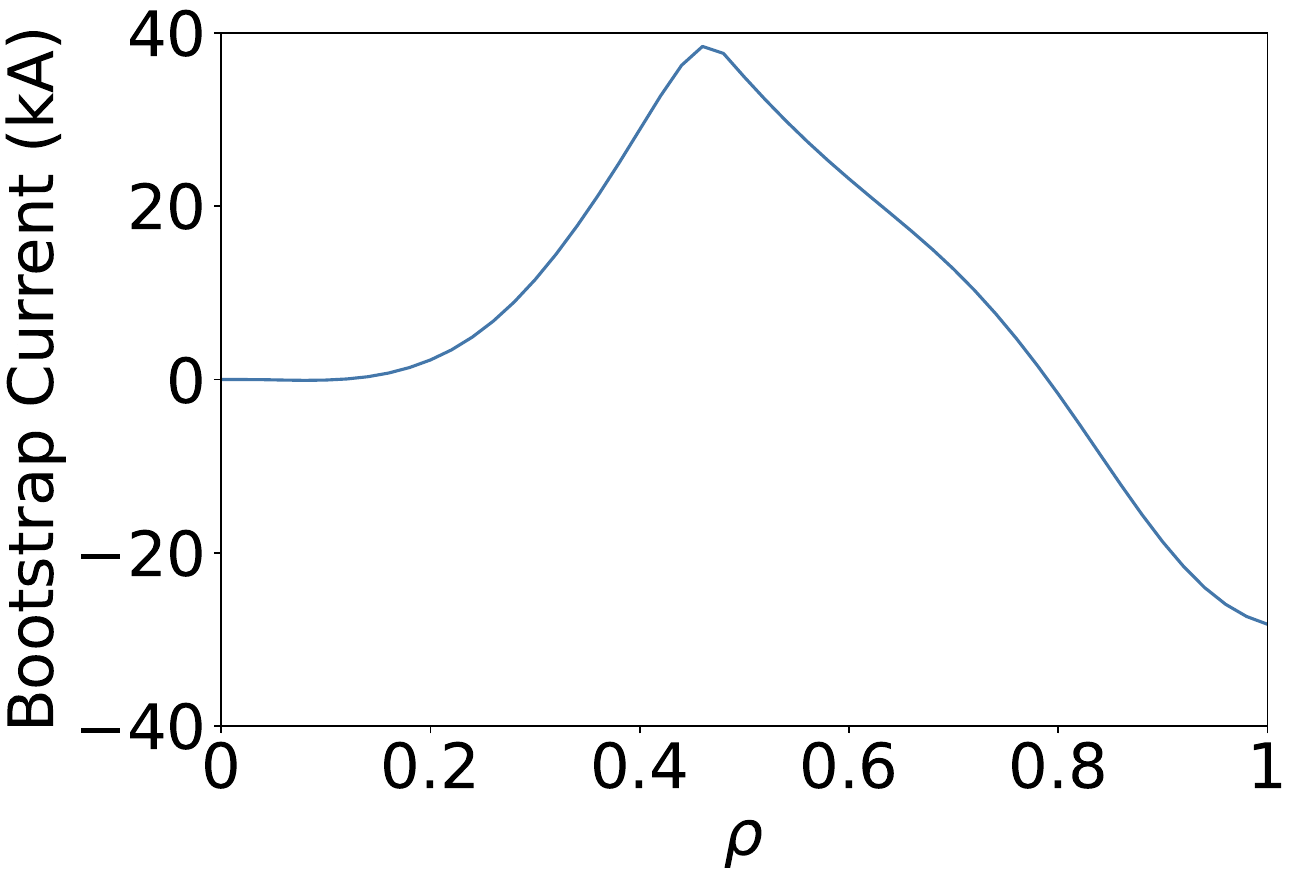}
    \caption{}
    \label{fig:Ibss}
    \end{subfigure}
    \begin{subfigure}[b]{0.43\textwidth}
    \includegraphics[width=\linewidth]{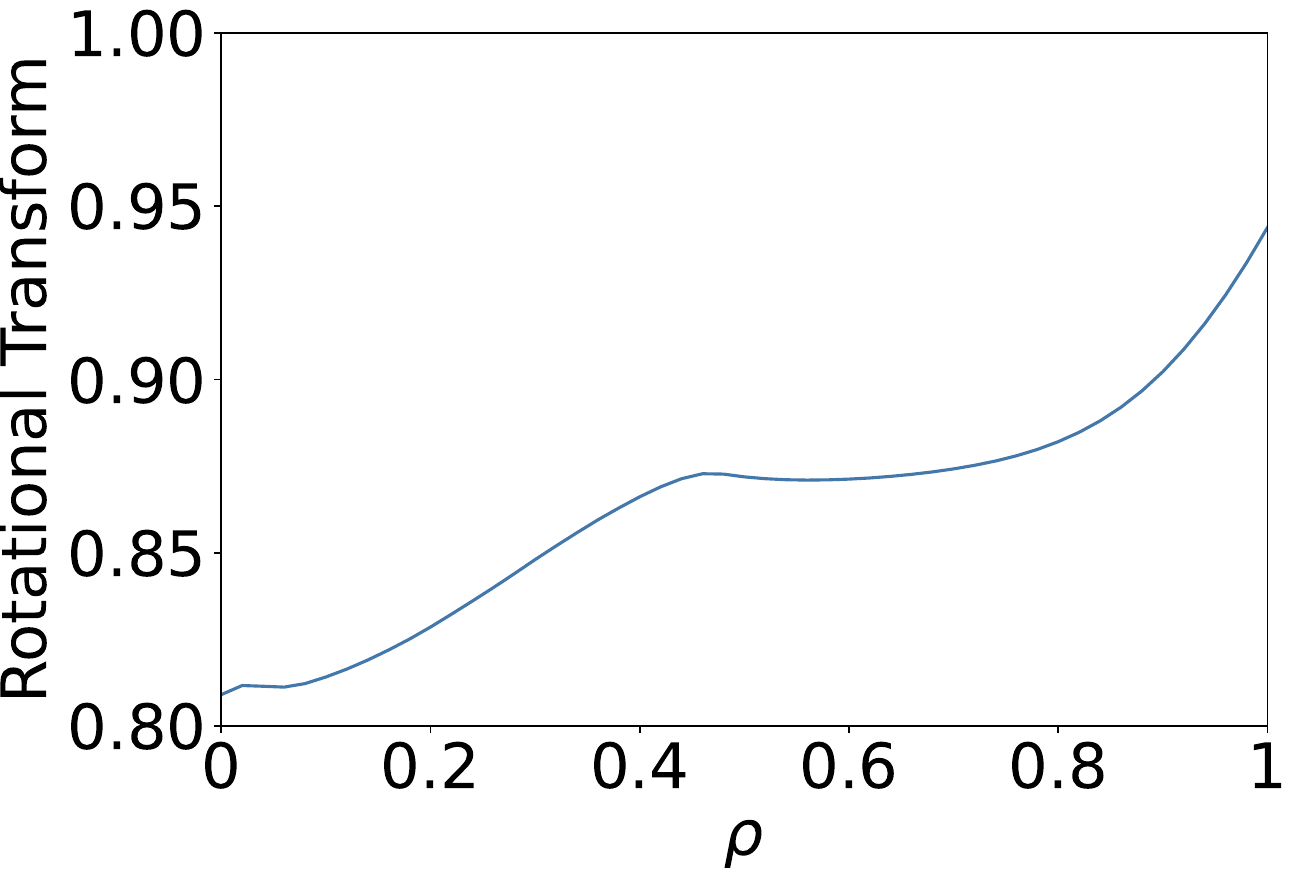}
    \caption{}
    \label{fig:iotas}
    \end{subfigure}
    \caption{NTSS results for the steady-state density (\subref{fig:lee3_n}), temperature (\subref{fig:lee3_T}), radial electric field (\subref{fig:Ers}), $\eiRat$ (\subref{fig:L11Rats}), bootstrap current (\subref{fig:Ibss}), and $\rotTrans$ (\subref{fig:iotas}) profiles of the example configuration.}
    \label{fig:ntss}
\end{figure}

\begin{table}
    \centering
    \caption{Plasma parameters of the example configuration. The mirror ratio does not change throughout a simulation because it is determined by the magnetic geometry alone. The ``Initial $P_{\alpha}$'' and ``Initial $P_{Br}$'' entries refer to the alpha power and bremsstrahlung power, respectively, at the beginning of the NTSS simulation.  All other entries refer to the end of the NTSS simulation. $\tauE$ is the energy confinement time and $\issof$ is calculated using the ISS04 scaling expression.}
    \begin{tabular}{|c|c|c|c|}
        \hline
        Quantity & Value \\
        \hline
        Mirror Ratio (\%) & 26.9 \\
        $\left< \beta \right>_{\mathrm{Vol}}$ (\%) & 3.14 \\
        Initial $P_{\alpha}$ (MW) & 495.6 \\
        Final $P_{\alpha}$ (MW) & 601.5 \\
        Initial $P_{Br}$ (MW) & 33.5 \\
        Final $P_{Br}$ (MW) & 72.7 \\
        $\tauE$ (s) & $2.34=1.23\issof$ \\
        \hline
    \end{tabular}
    \label{tab:ntss}
\end{table}

Figure~\ref{fig:lpks} shows that the last closed flux surface of the example configuration is slightly more elongated than that of \mbox{W7-X} high-mirror, but thanks to the boundary elongation objective we employed, the difference is relatively small.
Figure~\ref{fig:D11s} shows that the optimization of fuel-ion-relevant $\DooS$ for the example configuration has driven them to be substantially lower than those for \mbox{W7-X} high-mirror. The example configuration has also achieved its target $\epseff = 1\%$ almost perfectly throughout the core, as shown in figure~\ref{fig:epseffs}.
Because the fuel-ion-relevant $\DooS$ are well-optimized and $\epseff \approx 1\%$, we expect $\eiRat > 1$ in the core.
Additionally, figures~\ref{fig:Bcont} and \ref{fig:Jcont} show that in the core, the example configuration is reasonably quasi-isodynamic and has contours of the second adiabatic invariant that decrease monotonically with increasing minor radius. Neither of these features are explicitly targeted during the optimization.
The former feature may appear because the optimization drives the configuration toward having a magnetic field geometry that minimizes the radial drift of particles in the $\sqrt{\coll}$ regime (fuel ions), which is to say that the behavior of the fuel ions is nearly omigenous. The initial configuration employed for the optimization is approximately quasi-isodynamic, so as the optimizer searches for ``nearby'' field geometries with omnigenous fuel ion behavior, it is very likely to find another approximately quasi-isodynamic configuration.
The latter feature may appear because omnigenity is necessary for the maximum-$J$ property to be present~\cite{rodriguez_maximum-j_2024}, so optimizing for omnigenous fuel ion behavior likely drove the configuration to be approximately maximum-$J$ by accident.

Figure~\ref{fig:ntss} shows that $\eiRat$ is slightly greater than 1 for $0.15 \lesssim \normFluxLabel \lesssim 0.45$. This configuration consequently has a weak ion root for $0 \lesssim \normFluxLabel \lesssim 0.05$ and an electron root for $0.05 \lesssim \normFluxLabel \lesssim 0.45$. The peak strength of $\absE$ in the electron root region is roughly $15~\mathrm{kV/m}$, which is slightly greater in magnitude than its peak strength in the ion root region. The shape of the electric field profile in the ion root region is similar to that of \mbox{W7-X} for the same values of $\normFluxLabel$.
The difficulty in achieving an electron root deep in the core (which was observed several times in our testing) can perhaps be explained by the fact that flux surface shapes always tend to become elliptical as $\normFluxLabel \rightarrow 0$ \cite{landreman_direct_2018}. Confinement (omnigenity) of electrons places constraints on the contours of $B_{\mathrm{min}}$, whereas confinement of fuel ions places constraints on the contours of $B_{\mathrm{max}}$ \cite{cary_omnigenity_1997}. If cross-sections are almost purely elliptical, as is the case near the magnetic axis, it is difficult to optimize the magnetic field such that the latter constraints are (approximately) respected while the former are violated, which is necessary for the production of an electron root. There is more freedom to shape the flux surfaces far from the axis, so this task becomes easier. In addition to an electron root in most of the core, the example configuration has a relatively small bootstrap current (although we did not target this explicitly), almost monotonically positive magnetic shear, a reasonable power balance, and an energy confinement time $1.23$ times larger than that predicted by the ISS04 scaling expression \cite{yamada_characterization_2005} (see table~\ref{tab:ntss}). 
It is, however, Mercier unstable, which is unsurprising since Mercier stability was not among the optimization objectives.

\section{Discussion} 
The previous example has shown that it is possible to separately optimize neoclassical electron and fuel ion transport by modifying stellarator magnetic geometry. Furthermore, this differential optimization may be exploited to obtain a steady-state electron root in the core of a stellarator reactor. However, we were unable to generate any completely satisfactory configurations despite substantial effort. 
The presented configuration, for instance, has a small ion root deep in the core and is Mercier unstable.
Another configuration we generated used an optimization recipe very similar to the presented one to produce stronger fuel-ion-relevant $\DooS$ suppression, but effects related to the slope of the $\DooS$ vs $\normColl$ curves (like those in figure~\ref{fig:D11s}) and the resonant electric field prevented an electron root from being realized. 
Another configuration was initially optimized for quasi-poloidal symmetry to encourage the development of the electron root by lowering the principle poloidal variation of $B$ \cite{kuczynski_self-consistent_2024}, and while it had a very strong electron root throughout the core, was Mercier stable \cite{mercier_necessary_1960}, and had very small bootstrap current, it was likely far too elongated to be reactor-relevant.
Yet another configuration was produced by including only the aspect ratio and neoclassical targets from this work in its optimization recipe; it developed an electron root throughout the core but required too much initial heating power to be viable. 
These trials suggest that electron root optimization is challenging and that, while the fundamental idea of optimizing the electron and fuel ion neoclassical transport properties separately is sound, further development of the objective function would improve its reliability.

A possible extension of this work involves coupling a modern, fast neoclassical transport code such as KNOSOS \cite{velasco_knosos_2020, velasco_fast_2021} to a modern optimization framework such as SIMSOPT \cite{landreman_simsopt_2021} and performing optimizations similar to those presented in section~\ref{sec:results}. This would likely avoid some of the numerical difficulties we encountered while attempting to improve the optimization and allow for consideration of additional neoclassical effects. Notably, KNOSOS includes the tangential magnetic drifts (unlike DKES), which can be substantial for some configurations \cite{fujita_study_2021} and alter the confinement of fuel and impurity ions \cite{velasco_robust_2023}. The tangential drifts can also make the electron root more difficult to achieve in maximum-$J$ configurations \cite{calvo_effect_2017}. Because the maximum-$J$ property helps mitigate turbulence \cite{proll_turbulence_2022}, it may be a useful optimization target for future devices. A code that takes the tangential drifts into account would therefore be helpful if an electron root is to be targeted as well. Due to its speed, KNOSOS can also calculate thermal transport coefficients very quickly.
We performed some optimizations that used SFINCS \cite{landreman_comparison_2014} to target thermal transport coefficients as a ``polishing'' step after optimizing the monoenergetic transport coefficients. This was rather arduous but improved the neoclassical properties of the configurations, which suggests that targeting thermal transport coefficients throughout the optimization may be helpful --- Lascas Neto and coworkers \cite{neto_electron_2024} have recently used SFINCS to do so and obtained promising results.

Another possible extension of this work involves developing a more theoretical approach to ensuring excellent neoclassical confinement of fuel ions and moderate confinement of electrons. 
Helander and coworkers \cite{helander_optimised_2024} exploited the fact that in the formulation of omnigenity, there are separate conditions for zero average radial drift of marginally trapped particles (which pertains to fuel ions) and deeply trapped particles (which pertains to electrons). They derive an expression that allows the former conditions to be satisfied while the latter are violated and optimize a stellarator configuration using this expression as an objective. We suggest that further refinement of their analytical proxy, or perhaps a blend of the analytical and computational approaches, may provide the most reliable results for future stellarator optimization efforts.

Once a more reliable algorithm for electron root optimization has been devised, it will likely be beneficial to 
include a reasonably sophisticated turbulence model in the post-optimization transport simulations. This will indicate whether the large $\absE'$ in the core reduces turbulent fluxes. Recently devised simulation frameworks \cite{navarro_first-principles_2022} should provide this capability. Exploring the Pareto fronts between the electron root objective and other physics and engineering objectives will also be helpful for future design studies.

\section{Conclusions}
We have shown by example that neoclassical electron and fuel ion transport can be optimized separately in stellarators to raise the ratio of thermal transport coefficients $\eiRat$ and produce an outward-pointing ambipolar electric field (``electron root'') in the core under reactor-relevant conditions. Configurations with an electron root are expected to expel heavy impurities from the core during steady-state operation. We control $\Loo{e}$ through $\epseff$ and $\Loo{i}$ through an array of fuel-ion-relevant monoenergetic transport coefficients ($\DooS$); both $\epseff$ and $\DooS$ are characteristic of a given magnetic field and do not depend on plasma profiles. While our results are encouraging, further development is likely needed before the presented algorithm is ready for mainstream use.

\section*{Acknowledgments}
We thank Stefan Buller, Jos\'e Luis Velasco, Matt Landreman, Per Helander, Michael Drevlak, Alan Goodman, Eduardo Lascas Neto, Eduardo Rodr\'iguez, and Katia Camacho Mata for their insights and assistance. We also thank the anonymous referees, whose thoughtful feedback substantially improved this work.

This work was supported by a Fulbright grant from the German-American Fulbright Commission. 
This material is based upon work supported by the U.S. Department of Energy, Office of Science, Office of Advanced Scientific Computing Research, under Award Number DE-SC0024386.
This work has been carried out within the framework of the EUROfusion Consortium, funded by the European Union via the Euratom Research and Training Programme (Grant Agreement No.{\ }101052200 -- EUROfusion). Views and opinions expressed are however those of the author(s) only and do not necessarily reflect those of the European Union or the European Commission. Neither the European Union nor the European Commission can be held responsible for them.
Computations were performed on the Cobra and Raven HPC systems at the Max Planck Computing and Data Facility.

\newpage
\bibliographystyle{unsrt}
\bibliography{references}

\end{document}